\documentclass{INTERSPEECH2023}
\usepackage{subcaption}

\interspeechcameraready

\title{EE-TTS: Emphatic Expressive TTS with Linguistic Information}
\name{Yi Zhong$^1$, Chen Zhang$^1{^,}^2$, Xule Liu$^1$, Chenxi Sun$^1$, Weishan Deng$^1$, Haifeng Hu$^1$, Zhongqian Sun$^1$}
\address{
  $^1$Tencent AI Lab, China\\
  $^2$Zhejiang University, China
  }
\email{terryizhong@tencent.com, zc99@zju.edu.cn, 
\{xuleliu,baronsun,weishandeng,simbahu,sallensun\}@tencent.com}

\begin{document}

\maketitle
 
\begin{abstract}
While Current TTS systems perform well in synthesizing high-quality speech, producing highly expressive speech remains a challenge. Emphasis, as a critical factor in determining the expressiveness of speech, has attracted more attention nowadays. Previous works usually enhance the emphasis by adding intermediate features, but they can not guarantee the overall expressiveness of the speech. To resolve this matter, we propose Emphatic Expressive TTS (EE-TTS), which leverages multi-level linguistic information from syntax and semantics. EE-TTS contains an emphasis predictor that can identify appropriate emphasis positions from text and a conditioned acoustic model to synthesize expressive speech with emphasis and linguistic information. Experimental results indicate that EE-TTS outperforms baseline with MOS improvements of 0.49 and 0.67 in expressiveness and naturalness. EE-TTS also shows strong generalization across different datasets according to AB test results.
\end{abstract}

\section{Introduction}

Over the past few years, text-to-speech (TTS) models~\cite{tan2021survey,wang2017tacotron,ren2020fastspeech,kim2021conditional,zhang2021denoispeech} have made significant strides in enhancing intelligibility, quality, naturalness, and data efficiency. 
However, when it comes to some scenarios that require highly expressive speech, such as gaming dubbing and live streaming, original TTS systems may still generate speech with flat and mediocre prosody~\cite{kenter2020improving}, resulting in a lack of emotional resonance. Thus, researchers are paying more attention to expressive TTS, and have attempted to enhance speech expressiveness by improving the overall prosody~\cite{hayashi2019pre,xu2021improving, li2021improving, chen2021speech} or by capturing and modeling the emotion of speech~\cite{wu2022self,yoon2022language}.

Emphasis plays a vital role in determining speech expressiveness by affecting complex variations of many aspects of speech prosody, including pitch, phoneme duration, and spectral energy~\cite{eriksson2015acoustics,eriksson2018acoustic,eriksson2020lexical}. 
Thus, several studies have proposed emphasizing control in TTS systems~\cite{li2018emphatic,seshadri2021emphasis,liu2021controllable,shechtman2021supervised,stephenson2022bert}.
~\cite{li2018emphatic} enables emphasis control in an HMM-based TTS system by combining several handcrafted features. 
~\cite{seshadri2021emphasis, shechtman2021supervised} improve the emphasis effect by incorporating more intermediate acoustic features like pitch range \cite{shechtman2021supervised} or variance-based features \cite{seshadri2021emphasis}. 
However, the overall speech expressiveness of these works is still far from the ground truth in some scenarios that need highly expressive speech. Furthermore, they are unable to synthesize emphatic expressive speech without emphasis labels, which can be expensive to obtain during inference.
Therefore, \cite{stephenson2022bert} tried to predict emphasis position from the input text, while it still did not present a consistent controllability of emphasis and did not show a satisfied overall expressiveness of speech combined with the TTS model based on their results. 
According to some linguistics works ~\cite{beltrama2019conveying,xu2005phonetic}, the position and expression of emphasis highly depend on the syntax and semantics of the text. So one main reason for the above flaws from previous works is they do not consider the underlying principle or human inductive bias of emphasis like syntactic and semantic information to help the TTS model learn the distribution of expressive datasets.

In this paper, we propose Emphatic Expressive TTS (EE-TTS), a novel TTS model that utilizes linguistic information from syntax and semantics to generate emphatic expressive speech without emphasis labels. 
By incorporating two types of syntactic information, namely intra-word (the \underline{P}art-\underline{O}f-\underline{S}peech (POS) of each word) and inter-word (the \underline{D}ependency \underline{P}arsing (DP) features) as well as semantic information extracted through the pre-trained BERT~\cite{devlin2018bert} model, EE-TTS fully exploits linguistic information.
EE-TTS consists of 1) a linguistic information extractor to extract the syntactic and semantic information from the text; 2) an emphasis predictor to predict the positions of emphasis according to linguistic information; and 3) a conditioned acoustic model to generate expressive speech conditioned on emphasis positions and linguistic information. 
Besides, we use conformer~\cite{gulati2020conformer} instead of transformer as the encoder of the acoustic model due to its relative position embedding, which is corresponding to the importance of the relative position on emphasis~\cite{xu2005phonetic}.
Given the high cost of annotating emphasis labels for text, we take advantage of massive speech-text data without emphasis labels to pre-train the emphasis predictor and acoustic model by generating emphasis labels through a signal-based method. 

We conduct experiments on two mandarin TTS datasets with high expressiveness. The results show that EE-TTS can produce more expressive and natural speech with appropriate emphasis, surpassing current emphatic TTS systems by 0.49 of expressiveness and 0.67 of naturalness MOS improvements. We further carried out ablation studies to reveal the effectiveness of each aspect of linguistic information, the chosen architecture, as well as the pre-trained emphasis predictor.
To summarize, we outline the main contributions of this work as follows:

\begin{itemize}
\item By fully exploiting linguistic information (syntax and semantics), EE-TTS can predict more reasonable emphasis positions from the text.
\item Conditioned on the appropriate emphasis position and linguistic information, EE-TTS can consistently synthesize more expressive and natural speech with emphasis position.
\item High robustness and great generalization ability of EE-TTS are demonstrated according to experimental results\footnote{Some samples of synthesized speech for reference: 
\url{https://expressive-emphatic-ttsdemo.github.io/}}.

 \end{itemize}


\begin{figure*}[!thb]
	\centering
	\begin{subfigure}[h]{0.35\textwidth}
	    \captionsetup{justification=centering}
		\centering
		\includegraphics[width=\textwidth,trim={0.1cm 1cm 17cm 0cm}, clip=true]{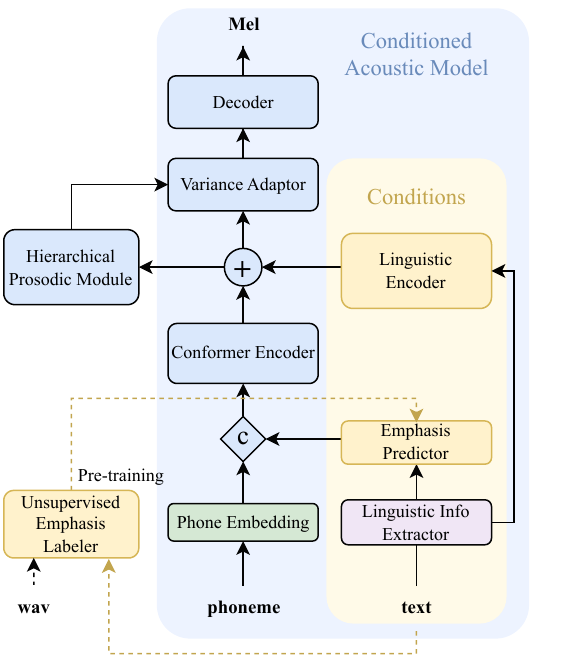}
		\caption{EE-TTS}
		\label{fig:f1_a}
	\end{subfigure}
	\begin{subfigure}[h]{0.29\textwidth}
	    \captionsetup{justification=centering}
		\centering
		\includegraphics[width=\textwidth,trim={10cm 1cm 8.6cm 0cm}, clip=true]{main_fig_v2.pdf}
		\caption{Linguistic Encoder}
		\label{fig:f1_b}
	\end{subfigure}
    \hspace{-5.2pt}
	\begin{subfigure}[h]{0.31\textwidth}
	    \captionsetup{justification=centering}
		\centering
		\includegraphics[width=\textwidth,trim={18cm 0.93cm 0.1cm 0cm}, clip=true]{main_fig_v2.pdf}
		\caption{Emphasis Predictor}
		\label{fig:f1_c}
	\end{subfigure}
	\caption{The entire framework of EE-TTS. The dashed lines in subfigure (a) indicate the pre-training procedure. Subfigures (b) and (c) show the detailed structure of the linguistic encoder and emphasis predictor respectively, as well as the linguistic information extractor.}
	\label{fig:f1}
	\vspace{-0.5cm}
\end{figure*}


\section{Proposed Method}

\subsection{Overview}
The overall architecture of EE-TTS is shown in Figure~\ref{fig:f1_a}.
We choose FastSpeech2~\cite{ren2020fastspeech} as the base architecture of the acoustic model, leveraging emphasis positions and linguistic embedding as conditions. These conditions are obtained through the emphasis predictor and the linguistic encoder respectively. 
The linguistic information extractor generates syntactic and semantic information from the input text, which is further fed into both the emphasis predictor and the linguistic encoder.
Inspired by~\cite{shechtman2021supervised}, a hierarchical prosodic module is incorporated to model nuanced prosody in speech. 

\subsection{Linguistic Information Extractor}
To involve inductive bias in linguistics, we propose linguistic information extractor, as shown in the purple blocks in Figure~\ref{fig:f1_b} and \ref{fig:f1_c}.
To extract syntactic information, we first use jieba\footnote{\url{https://github.com/fxsjy/jieba}} to segment the input text and then use pyltp\footnote{\url{https://pypi.org/project/pyltp/}} to predict Part-of-Speech (POS) tags and Dependency Parsing (DP) relations of all words at the intra-word and inter-word levels respectively. 
The DP result is presented graphically, with each word having only one out edge except the root words. 
As for semantic information, we use pre-trained BERT~\cite{devlin2018bert} is used to capture it at the character level. 
The multi-level linguistic information also reflects the hierarchical structure of the text~\cite{brolin2017hierarchy}. 

\subsection{Conditioned Acoustic Model}
The acoustic model synthesized the speech conditioned on emphasis positions and linguistic embedding. 
By combining Convolutional neural network and Transformer, Conformer~\cite{gulati2020conformer} can model both local and global dependencies, showing impressive performance in various tasks.
Thus, we choose Conformer as the encoder for EE-TTS to benefit from its ability to model hierarchical structure.
Moreover, Conformer incorporates relative positional encoding from TransformerXL~\cite{dai2019transformer}, to handle input sequences of varying lengths and generate more accurate position information.

Both emphasis positions and linguistic embedding are generated from extracted linguistic information.
The emphasis positions are given by the emphasis predictor, which is described in Section~\ref{sec:emphasis_predictor}.
Figure~\ref{fig:f1_b} illustrates the process of obtaining linguistic embedding. DP relations are first serialized by a Graph2Relation operation, which selects the type of the only one out edge of each word as the label and assigns root labels for the root words. 
DP relations and POS tags undergo two separate embedding layers and are then expanded with the output BERT through a phone-level length regulator to keep consistent with the size of the encoder output. 
These three are summed to a linguistic embedding and added to the encoder outputs.


\subsection{Emphasis Predictor}
\label{sec:emphasis_predictor}
Figure~\ref{fig:f1_c} depicts the details of the emphasis predictor.
To generate the initial vector for each character node, we first embed the intra-word POS tags and expanded them to the character level. After that, we concatenate them with the character-level outputs of BERT.
For the DP relations, we add the BOS (Begin Of Sentence) and EOS (End of Sentence) nodes in the DP relation graph and then encode the graph with the node initial vector using a Gated Graph Neural Network (GGN)~\cite{li2015gated} to obtain the character-level features. 
Finally, two linear layers are employed to predict the binary classification result for each character. A value of 0 denotes the absence of emphasis on that character, while a value of 1 indicates the presence of emphasis. The character-level labels are then passed through an embedding layer and expanded to the phone level to concatenate with the phone embedding.

 \subsection{Pre-train with unsupervised emphasis labeling}
Pre-train and finetune paradigm is quite common these days to benefit from pre-trained big models on large-scale datasets\cite{chen2018sample,chen2021adaspeech}. 
However, collecting a vast amount of labeled data on emphasis is extremely challenging and expensive due to the significant subjectivity involved in determining whether a word is emphasized or not, and the ambiguity that a single sentence may have more than one valid emphasis pattern. 
To make use of unlabeled data in a cost-effective way, we employ the Wavelet Prosody Toolkit\footnote{\url{https://github.com/asuni/wavelet_prosody_toolkit}} \cite{suni2017hierarchical} to obtain pseudo emphasis labels. This toolkit is based on the continuous wavelet transform (CWT) of a weighted sum of the pitch, energy, and duration signals to calculate a prominence score for each character. These scores are quantized into two categories to indicate whether a character is emphasized or not for pre-trained datasets. Both the acoustic model and the emphasis predictor are pre-trained with these pseudo-emphasis labels.

\section{Experiments}

\subsection{Dataset}
\subsubsection{TTS Training Data}
For pre-training, we used a Mandarin dataset consisting of approximately 90,000 utterances with a total length of approximately 80 hours. The dataset includes several single-speaker datasets such as an open-source female corpus from data-baker \cite{baker2017chinese} and a multi-speaker and multi-style dataset of approximately 30 hours, which includes 60 speakers and 7 different speech styles. All datasets were trimmed to remove silence at the beginning and end and were downsampled to 24 kHz. The dataset was randomly divided into a training set of 89,000 sentences and an evaluation set of 1,000 sentences. 

For fine-tuning, we used a private Mandarin dataset in gaming style with a total of 3,500 utterances from a female speaker (F1). About 1,800 utterances and 3,000 characters in the dataset are emphasized. To evaluate the model generalization ability to other expressive datasets, we also finetuned another dataset in live streaming style with a total of utterances 3,000 from another female speaker (F2), and about 1,200 utterances and 2,200 characters are
emphasized. The emphasis positions of these two datasets are labeled by one professional annotator with sufficient training. 
\subsubsection{Emphasis Predictor Training Data}
We use the same pre-training dataset with the TTS model for position predictor, a total of 90,000 sentences with unsupervised emphasis labels. For finetuning, we utilized the emphasis position confidence scores of the pre-trained model to filter the unsupervised emphasis labels and selected a total of 9,550 sentences with high confidence. These sentences combined with 6,500 sentences from speakers F1 and F2 to be our fine-tuning dataset of emphasis predictor. 
\subsection{Model Configurations}
\subsubsection{TTS Training Configurations}
We utilize the basic configuration of the Fastspeech2 \cite{ren2020fastspeech} for the models listed below unless otherwise explained. We choose the FastSpeech2 with an emphasis embedding and the hierarchical prosodic module \cite{shechtman2021supervised} as the baseline to compare fairly. For our proposed model, the conformer encoder has 4 layers with both input  and encoder dimensions of 256 and 2 attention heads, following the implementation and the default configurations of Espnet\footnote{\url{https://github.com/espnet/espnet}}. We use \textit{bert-base-chinese} available on HuggingFace\footnote{\url{https://huggingface.co/bert-base-chinese}} as our pre-trained BERT and fine-tuned with our model training. We trained all models to 250,000 steps on 4 Tesla V100-16GB GPUs with batch size 64. We modified the first anneal step to 200,000 due to the size of the fine-tuning dataset. The baseline model is trained to 250,000 steps only with the fine-tuning dataset. For all the other models, we first pre-trained the model to 180,000 steps with the pre-trained dataset and then fine-tuned it to 250,000 steps. For the two models with BERT, we use an independent learning rate with exponential decay of 0.7 rather than 0.5 for FastSpeech2 to prevent the BERT module from not converging. Besides, we use a default HiFiGAN~\cite{kong2020hifi} trained with the fine-tuning dataset as the vocoder for all TTS models.

\subsubsection{Emphasis Predictor Configurations}
For the emphasis predictor, we use the same pre-trained BERT model in the TTS system and fine-tuned it with the model. The POS embedding size is set to 30. The GGN module is implemented following the default configuration with a total of 17 relations including BOS and EOS and 3 iterations, the output size is 512 following two linear layers with 128 and 2 output sizes. Our pre-training and fine-tuning processes were both trained for 50 epochs with a learning rate of 5e-5 on a Tesla V100-16GB GPU with batch size 32.
\section{Results}
\subsection{Evaluation Metrics}

We conduct several subject tests to evaluate our model comprehensively. A total of 180 utterances are randomly shuffled for MOS tests. The overall naturalness and expressiveness of speech are evaluated by a standard naturalness MOS (N-MOS) and an expressiveness MOS (E-MOS). For both MOS tests, 25 native raters read the descriptions and listened to the audio examples of each level first, then listened to all the utterances and gave 5 scaled scores from 1 to 5 based on their subjective perception of how naturalness or expressiveness of each speech. 
Apart from the subjective tests, we also evaluate a commonly used objective metric called Root Mean Square Error (RMSE) of Fundamental Frequency (F0) for Ablation Studies in Section 4.2. We calculate the average RMSE of the F0 in Hertz of random 100 utterances in the validation set as an auxiliary metric to evaluate which model fit the pitch prosody better. We also performed AB preference tests by 15 listeners resulting in a total of 100 utterances to verify the generalization to different datasets conveniently. All the tests listed below are done in speaker F1, except the AB preference tests are done for both F1 and F2.

\subsection{Overall Performance Evaluation}
\begin{table}[htbp]
\centering
\caption{Results of Naturalness and Expressiveness MOS for different TTS systems with 95\% confidence intervals.}
\begin{tabular}{lcc}
\hline
Method & N-MOS & E-MOS \\
\hline
\textit{Ground Truth}\hspace{6mm} & 4.66 $\pm$ 0.05 & 4.65 $\pm$ 0.05 \\ \hline
\textit{Baseline (GT)}\hspace{6mm} & 3.67 $\pm$ 0.06 & 3.76 $\pm$ 0.06\\
\textit{Proposed (GT)}\hspace{6mm} & 4.34 $\pm$ 0.06 & \textbf{4.25 $\pm$ 0.06}\\
\textit{Proposed (Pred)} \hspace{6mm} & \textbf{4.37 $\pm$ 0.06}& 4.24 $\pm$ 0.06\\
\hline
\end{tabular}
\label{tab:results}
\end{table}

In table \ref{tab:results}, we compare the naturalness MOS and expressiveness MOS for Ground Truth (GT), our proposed model and baseline model with the human-labeled emphasis position of GT. We also provide the result of the proposed model with the labels predicted from our emphasis predictor. It is apparent that our proposed model significantly outperforms the baseline in both naturalness and expressiveness (p-values$\ll$0.05).  The MOS results of the proposed model with predicted labels even slightly outperform the one with Ground Truth labels, which may be because the position predictor tends to predict the emphasized characters that are present in the training set, and at the same time, the TTS model learns better emphasis expression on these characters.
\begin{figure}[htp]
  \centering
  \includegraphics[width=\linewidth]{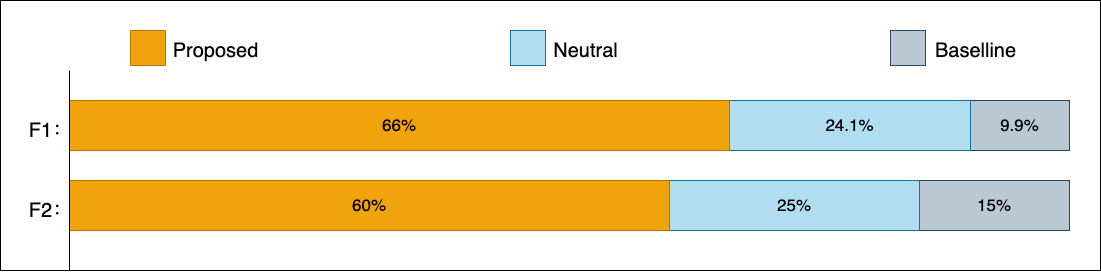}
  \caption{AB preference results between EE-TTS and baseline of two datasets.}
  \label{fig:ab}
\end{figure}

Fig \ref{fig:ab} shows the ab preference test results between EE-TTS and the baseline for two datasets, judgers prefer those speeches generated from the proposed model notably. A similar result between speaker F1 and F2 shows EE-TTS generalize to different datasets well.
In the AB preference tests of two datasets, approximately 10 percent of the baseline audio samples were favored more over the proposed models, as in this portion, the positions predicted by the emphasis predictor are not reasonable enough. Unreasonable emphasis positions can lead to a decrease in the human perception of naturalness and expressiveness though the overall prosody has more variation.

\subsection{Ablation Studies}
We conduct ablation studies step by step to evaluate the effectiveness of each module in the proposed model. A total of 6 settings are involved for comparison: 1) EE-TTS; 2) EE-TTS equipped with a transformer encoder, not conformer, denoted as \textit{-C+T}; Settings 3) 4) and 5) remove the BERT module, dependency parsing module, and part-of-speech module step by step, denoted as \textit{-BERT}, \textit{-DP}, and \textit{-POS} respectively; 6) EE-TTS without unsupervised labels (UL) pre-training procedure. 
\begin{table}[htbp]
\centering
\caption{Results of ablation studies of the acoustic model. C stands for Conformer and T stands for Transformer in the second setting. The last row shows the results of the EE-TTS without Unsupervised Labels(UL) during pre-training}
\begin{tabular}{lccc}
\hline
Method & N-MOS & E-MOS & F0-RMSE \\ \hline
\textit{EE-TTS} & \textbf{4.34 $\pm$ 0.06} & \textbf{4.25 $\pm$ 0.06} & \textbf{53.495} \\
 \textit{\hspace{3mm}-C+T} & 4.19  $\pm$ 0.06 & 4.12 $\pm$ 0.06 & 56.979 \\
 \textit{\hspace{6mm}-BERT} & 4.03 $\pm$ 0.06 & 4.01 $\pm$ 0.06 & 57.352 \\
 \textit{\hspace{9mm}-DP} & 4.11 $\pm$ 0.06 & 4.05 $\pm$ 0.06 & 57.405 \\
 \textit{\hspace{12mm}-POS} & 3.99 $\pm$ 0.07 & 4.01 $\pm$ 0.06 & 57.956 \\ \hline
  \textit{EE-TTS w/o UL} & 3.97 $\pm$ 0.06 & 4.04 $\pm$ 0.06 & 53.865 \\ \hline
\end{tabular}
\label{tab:results2}
\end{table}

As shown in Table \ref{tab:results2}, the BERT module and the conformer encoder helped improve both the expressiveness and naturalness of speech significantly (p-values$\ll$0.05), so did pre-training with unsupervised emphasis labels. The part-of-speech features helped only improve naturalness but did not benefit the expressiveness much, and the employment of dependency parsing seems to have no advantages. For the F0-RMSE, we can see this metric gradually decreases as each module is added step by step, and the trend matches the trend of the MOS score increase indicating the consistency of our results. 
However, the existence of part-of-speech and dependency parsing features do not affect the expressiveness much. It may be because these two types of information only affect where the emphasis appears but not how people express the emphasis acoustically, or due to the error propagation from the systems used to predict and extract them. 
We also observed a strong positive correlation between the MOS scores for expressiveness and naturalness in our results. This may owe to the fact that for audio with low naturalness, even if it has more variations, it may not resonate with the listener due to the lack of human likeness. 

Ablation studies for the emphasis predictor are conducted to further reveal the effectiveness of linguistic information. The precision, recall, and F-score results of these methods are reported in Table \ref{tab:results3}.  The results indicated the benefits of the performance of emphasis prediction by leveraging each kind of linguistic information. Besides, we also give a metric called reasonable precision (R-Precision) to indicate the rate of predicted positions that are reasonable for human subjective judgment. Since the location of emphasis in the speech has high variability and individuality, in real-world applications, we care more about whether the predicted emphasis locations sound appropriate to humans rather than need them exactly located at the same position as the ground truth. Similar MOS results between our proposed model with predicted labels and GT labels in Table \ref{tab:results} also confirm this assumption.

\begin{table}[htbp]
\centering
\label{tab:my-table}
\caption{Results of ablation studies of the emphasis predictor}
\begin{tabular}{lcccc}
\hline
Methods & Precision & Recall & F-score & R-Precision \\ \hline
\textit{Proposed} &\textbf{0.52} & \textbf{0.63} & \textbf{0.57} & \textbf{0.87} \\ 
\textit{\hspace{2mm}-DP} &0.50 & 0.57 & 0.53 & 0.86 \\ 
\textit{\hspace{4mm}-POS} &0.52 & 0.53 & 0.53 & 0.84 \\ 
\textit{\hspace{6mm}-BERT}  &0.41 & 0.45 & 0.43 & 0.77 \\ \hline
\end{tabular}
\label{tab:results3}
\end{table}

\section{Conclusions}
In this paper, we proposed the Emphatic Expressive TTS (EE-TTS) model,  a novel approach that offers a promising solution for generating both expressively and naturally emphasized speech without emphasis labels. By leveraging linguistic information such as syntax and semantics, the model predicts more reasonable emphasis positions and produces more expressive emphasized speech compared to the baseline. Furthermore, the ablation studies demonstrate the integration of the BERT model and the conformer encoder allows the model to capture semantic as well as relative position information, resulting in significant improvements in both the expressiveness and naturalness MOS. Notably, the proposed model does not require input emphasis labels and it is generalizable shown in the AB test, making it easy to apply in practice.  This approach, though benefits from pre-training with unsupervised labels, still needs a certain amount of human-labeled emphasis labels for the fine-tuning dataset. In future research, our goal is to make the most of large-scale base data by enhancing the precision of unsupervised emphasis labeling tasks, which may allow us to eliminate the need for human labeling and ultimately improve the overall utilization of the data. We also look forward to further evaluate generalization capabilities in additional
languages. Overall, our work fully exploits linguistic information to generate highly expressive TTS with appropriate emphasis and paves the way for future research.

\ifinterspeechfinal

\else

\fi


\bibliographystyle{IEEEtran}
\bibliography{mybib}

\end{document}